\def\arxiv{1}

\ifnum\arxiv=1
  \documentclass{clear2027}
\else
  \documentclass[anon,12pt]{clear2027}
\fi
\ifnum\arxiv=1
  \jmlrvolume{}\jmlryear{}\jmlrpages{}\jmlrproceedings{}{}\jmlrworkshop{}\editor{}
\fi
\makeatletter\renewcommand*{\@titlefoot}{}\makeatother
\ifnum\arxiv=1
  \usepackage{placeins}
  \usepackage{etoolbox}
  \pretocmd{\section}{\FloatBarrier}{}{}   
  \makeatletter
  \renewcommand{\fps@figure}{!htbp}\renewcommand{\fps@table}{!htbp}
  \makeatother

\else
  \makeatletter
  \renewcommand{\fps@figure}{h}\renewcommand{\fps@table}{h}
  \makeatother
\fi
\usepackage{times}
\usepackage{booktabs}
\usepackage{graphicx}

\title[Doubly robust estimation of direct and spillover effects]{Doubly robust estimation of direct and spillover effects\\using multiple pre-treatment periods}

\clearauthor{%
 \Name{Martin Huber} \Email{martin.huber@unifr.ch}\\
 \addr University of Fribourg, Department of Economics\\
 Bd.\ de P\'{e}rolles 90, 1700 Fribourg, Switzerland%
}

\newtheorem{prop}{Proposition}
\newtheorem{assumption}{Assumption}

\newcommand{\Exp}[1]{E\!\left[#1\right]}
\newcommand{\I}{\mathrm{I}}
\newcommand{\cT}{\mathcal{T}}
\newcommand{\cN}[2]{\mathcal{N}^{#1}_{#2}}
\newcommand{\cTc}[2]{\mathcal{T}^{#1}_{#2}}

\begin{document}
\maketitle

\begin{abstract}%
Under interference, the treatment of some units affects the outcomes of others, so
that the effect on a treated unit consists of a direct and a spillover component. A
non-treated comparison group then generally identifies neither component, but the
difference between the total effect on the treated and the spillover effect on the
comparison units.  Under certain identifying assumptions, both components may be identified separately when multiple pre-treatment periods are available, even in the absence of an uncontaminated comparison group. Under a no-anticipation assumption, pre-treatment outcomes are free of both components for every unit. Extrapolating the pre-treatment trend within a group, as in interrupted time-series approaches, therefore identifies the counterfactual outcome in the absence of any treatment, provided that the model used for the extrapolation is correctly specified. Applied to the treated it gives the total effect on them, applied to the non-treated at a given level of exposure the spillover effect on them. Disentangling the two components in addition requires effect homogeneity across treated and non-treated units with the same exposure and covariates, as well as common support in exposure. We derive doubly robust and Neyman-orthogonal
expressions for the effects of interest, which permits controlling for possibly high-dimensional covariates by machine learning. A simulation study confirms the root-$n$ consistent behaviour of the estimators and the reliability of the standard errors under specific regularity conditions.
We apply the framework to the compulsory licensing of enemy-owned patents in the
United States in 1919. Licensing raised patenting in the licensed subclasses
and in technologically close ones alike, so that a comparison-group design
understates the total effect.%
\end{abstract}

\ifnum\arxiv=1 \vskip 2ex \fi


\ifnum\arxiv=1 \clearpage \fi

\section{Introduction}\label{sec:intro}

Most studies in causal analysis and treatment evaluation explicitly or implicitly
rule out interference or spillover effects, as implied by the stable unit
treatment value assumption (SUTVA), see \citet{Cox58} and \citet{Rubin80}, which
is frequently implausible in empirical applications. When treating some units
affects the outcomes of others, the effect of the treatment on a treated unit
consists of the direct effect of its own treatment and the spillover effect of the
treatment of others, which together form the total effect on the treated.
Moreover, a non-treated comparison group may itself be contaminated by spillovers,
in which case the estimand of a causal inference method is in general the
difference between the total effect on the treated and the spillover effect on
the comparison units, see \citet{MealliViviens2026}. Neither component is
identified separately, and the sign of the estimand is uninformative about the
sign of either. The remedies proposed so far restrict the interference structure
and require the existence, and the correct identification, of units that are clean
controls in the sense of being unaffected by any spillover.

In this paper, we consider the evaluation of total, direct, and spillover effects based on approaches that require no clean controls but outcomes observed in multiple pre-treatment periods, while also controlling for observed covariates. The underlying intuition is that under no anticipation of treatment, the pre-treatment outcomes of every unit, treated or not, are realizations of the potential outcome in the absence of any treatment, such that interference contaminates the post-treatment outcomes of the comparison group but not its pre-treatment outcomes. Extrapolating the pre-treatment trend within a group, as in interrupted time series (ITS) approaches, therefore delivers the counterfactual outcome in the absence of any treatment, provided that the extrapolation is valid. Applied to the treated, this identifies the total effect on the treated; applied to the non-treated at a given level of exposure to the treatment of others, the spillover effect on them. \citet{Cellietal2026} formally establish this identification result without covariates. With two pre-treatment periods and a linear trend the counterfactual conditional mean is a second difference of the cell's own pre-treatment conditional means, and further periods permit richer trend specifications. Neither a comparison group nor a common support condition across groups is required.  Our first contribution is a conditional
version of this approach, which permits trends to differ across observed characteristics and the covariate composition of a group to change across periods, and thereby extends the argument from panels to repeated cross sections.

Exposure is frequently summarized by an exposure mapping in the sense of \citet{AronowSamii2017}, a function of the treatments of all other units measuring how intensely a unit is reached by them, such as the number of treated units in its neighborhood. In this context, separating the total effect on the treatment into its direct and exposure-related spillover effects is demanding, as the difference between the two is the spillover treated units would experience in the absence of their own treatment, which is never observed. Our second contribution is to make precise what it takes to recover it. First, an effect homogeneity condition is required, namely that treated and non-treated units with the same exposure and covariates experience the same average spillover, which is restrictive whenever spillovers differ along unobserved characteristics in which the two groups differ. Second, exposure must be observed for both groups and have common support across them, which fails when treatment status mechanically
determines exposure. Where either is implausible, the analysis is confined to the total effect on the treated and the spillover effect on the non-treated, which we regard as the main parameters.

Our third contribution is a testable restriction. When some units are classified as
non-exposed, their spillover is zero by construction, so the corresponding estimate
must not differ significantly from zero. A rejection indicates either that the
identifying assumptions fail or that units classified as non-exposed are in fact
reached by the treatment. The test thus assesses the design jointly with the
exposure mapping. It requires the mapping to claim that some units are non-exposed,
but not that this claim be correct. We state it conditional on covariates, because
spillovers of opposite sign in different regions of the covariate space cancel upon
averaging. A second restriction arises if conditional parallel trends are imposed in
addition: the counterfactual is then overidentified, and the two routes agree if
and only if the conditional pre-treatment trends coincide across groups, which is
what placebo tests in difference-in-differences analysis assess.

Finally, we derive doubly robust and Neyman-orthogonal expressions for the effects
of interest, permitting estimation of the conditional means and propensity scores
by machine learning, and establish root-$n$ consistency and asymptotic normality. A
simulation study shows the estimators to be close to unbiased, to converge at the
root-$n$ rate and to have reliable standard errors when the assumptions hold, and
the overidentification test to have power against violations of conditional parallel
trends. In an application to the compulsory licensing of enemy-owned patents in the
United States in 1919 previously analyzed by \citet{MoserVoena12}, we find that licensing raised patenting both in the licensed
subclasses and in technologically close ones.

\citet{MealliViviens2026} show that comparison-group methods identify a difference between the total effect on the treated and the spillover on the non-treated under unrestricted interference. They also show that within-group before-and-after comparisons separate the two components under a constant-mean restriction, which we replace by a trend model. Identification of direct and spillover effects under a known exposure mapping is studied in difference-in-differences and related panel designs by \citet{clarke2017did}, \citet{butts2021difference}, \citet{HuberSteinmayr2021}, \citet{xu2023differenceindifferences}, \citet{Hettingeretal2024} and \citet{SunXiao2025}. Cross-sectional designs under randomization or selection on observables are considered by \citet{HudgensHalloran2008}, \citet{TchetgenTchetgenVanderWeele2012B}, \citet{Forastiereetal2021} and \citet{leung2022unconfoundedness}. In all of these, effects defined relative to the absence of any treatment are identified through comparison units the treatment does not reach. This also applies to the synthetic control approaches with spillovers of \citet{Grossietal2023} and \citet{distefano2024inclusive}.

Counterfactual forecasting without a comparison group goes back to the intervention analysis of \citet{BoxTiao1975} and is developed by \citet{Menchettietal2022}, \citet{cerqua2022losing} and \citet{botosaru2023forecasted}. \citet{nutarelli2026assessing}, for instance, apply such methods to assess the total effect of the COVID-19 shock on exporting firms. In their conclusion, they informally point out that these methods may also be used to investigate spillover effects, since they require no contemporaneous control group of untreated units. To the best of our knowledge, \citet{Cellietal2026} are the first to formally analyze counterfactual forecasting under interference, and their paper is the closest to ours. They show that under specific conditions, counterfactual forecasting from pre-treatment outcomes identifies the total effect on the treated and the spillover effect on non-treated units at a given level of exposure, even if every non-treated unit is affected by the treatment. They also characterize which kinds of effects are identified when exposure to spillovers is latent, that is, unobserved for each unit, or when spillovers affect all units. We extend their identification results to condition on covariates, which also covers repeated cross sections, and derive doubly robust estimators. Furthermore, we provide conditions for decomposing the total effect on the treated into its direct and spillover components. We also propose tests of specific identifying assumptions, controlling for covariates based on the doubly robust approach of \citet{Apfel2023learning}.

The remainder of the paper is organized as follows. Sections~\ref{sec:effects} to \ref{sec:test} present the effects of interest, the assumptions, identification, doubly robust estimation
and the testable restrictions, Sections~\ref{sec:sim} and \ref{sec:app} the
simulation study and the empirical application, and Section~\ref{sec:conclusion}
concludes.

\section{Causal effects of interest}\label{sec:effects}

We consider $n$ units observed in $K+1\geq2$ pre-treatment periods
$t\in\{-K,\dots,-1,0\}$ and one post-treatment period $t=1$, and denote by $Y_t$ the
outcome and by $D\in\{0,1\}$ a unit's own treatment, which is realized between
$t=0$ and $t=1$. Let
$\mathbf{D}_{-i}$ be the treatment vector of all units other than $i$. Rather than
restricting the interference structure, we summarize it by an exposure mapping in
the sense of \citet{AronowSamii2017}, that is, a function
\begin{align}\label{expmap}
G_i=g(\mathbf{D}_{-i},F_i)\in\mathcal{G},
\end{align}
which maps the treatments of all other units into a low-dimensional measure of how
intensely unit $i$ is reached by them. Here $F_i$ collects the characteristics
determining which units can reach $i$, such as spatial coordinates or network
positions, and $\mathcal{G}$ is a finite set containing the element $0$, which
denotes the absence of exposure. Leading examples are the share of treated households
in a village or an indicator for having a treated neighbour in a network. Potential outcomes in the sense of \citet{Neyman23} and
\citet{Rubin74} are indexed by own treatment and exposure, $Y_t(d,g)$, which
replaces the no-interference part of SUTVA by the weaker restriction that $G$ is a
sufficient summary of $\mathbf{D}_{-i}$ for unit $i$'s outcome. See
\citet{HalloranStruchiner1991} and \citet{Manski2013} for the general formulation
of interference. The observed post-treatment outcome is $Y_1=Y_1(D,G)$. We also
observe covariates $X$, which may be time-varying, in which case $X$ collects the
values across the periods considered. Extrapolating the counterfactual outcome is a
forecasting problem, in which post-treatment covariate values such as seasonality or
local economic conditions are potential predictors. The requirement is that $X$ be
unaffected by $D$ and $G$, which is innocuous for pre-treatment values but needs
scrutiny for post-treatment ones, as discussed without interference by
\citet{caetano2022difference}. Interference enlarges the set of potentially affected
covariates, since a post-treatment covariate may respond to the treatment of others
as well as to a unit's own. 

Since interference generally rules out comparing a unit's outcome to a world in
which only that unit is non-treated, the effects are defined relative to the
benchmark in which no unit is treated. We define all parameters conditional on the covariates
and obtain the aggregates by averaging over the respective subpopulation observed in
the post-treatment period; we suppress $T=1$ in the definitions and make it explicit
in the identification results. The total effect on the treated,
\begin{align}\label{tatt}
\tau_{T}^{1}(x)=\Exp{Y_1(1,G)-Y_1(0,0)\mid X=x,D=1},\qquad
\tau_{T}^{1}=\Exp{\tau_T^{1}(X)\mid D=1},
\end{align}
comprises the direct effect of own treatment and the interference, or spillover,
effect of the treatment of others. We write $\tau_D$ for direct and $\tau_I$ for
interference effects, superscripted $1$ for effects on the treated and $0$ for
effects on the non-treated. The interference effect on the non-treated with
exposure $g$ is
\begin{align}\label{se}
\tau_{I}^{0}(x,g)=\Exp{Y_1(0,g)-Y_1(0,0)\mid X=x,D=0,G=g},
\end{align}
with aggregate $\tau_{I}^{0}(g)=\Exp{\tau_I^{0}(X,g)\mid D=0,G=g}$. Its argument $g$
is the exposure level at which the spillover is evaluated.

Among the treated, a direct effect switches own treatment at a fixed exposure
level and an interference effect switches exposure at a fixed level of own
treatment,
\begin{align}\label{de}
\tau_{D}^{1}(g)=\Exp{Y_1(1,g)-Y_1(0,g)\mid D=1},\qquad
\tau_{I}^{1}(d)=\Exp{Y_1(d,G)-Y_1(d,0)\mid D=1},
\end{align}
where the argument of each effect is the level at which the variable not switched is
held, the switched one moving between its realized value and zero. The total effect then admits the two
decompositions
\begin{align}\label{pure}
\tau_T^{1}=\tau_{D}^{1}(G)+\tau_{I}^{1}(0)=\tau_{D}^{1}(0)+\tau_{I}^{1}(1),
\end{align}
which differ in the level at which the argument not being switched is held, and
which coincide when own treatment and exposure do not interact, the interaction
being $\tau_D^1(G)-\tau_D^1(0)=\tau_I^1(1)-\tau_I^1(0)$. Which decomposition is of
interest depends on the policy question. $\tau_D^1(G)$ is the effect of treating a
unit in the realized configuration of the program, whereas $\tau_D^1(0)$ is the
effect of treating it in isolation, which is the relevant quantity when the program
is to be scaled down or implemented elsewhere. $\tau_I^1(1)$ measures how much of
the total effect is due to the treatment of others.

Throughout, we work with conditional mean outcomes rather than with
unit-level differences, which keeps the analysis applicable to repeated cross
sections, where a within-unit difference across periods cannot be formed. Let $T$
index the period, let a cell $\mathcal{C}$ be a subpopulation defined by a value of
$(D,G)$, and write $\mu_t(x,\mathcal{C})=\Exp{Y_t\mid X=x,\mathcal{C},T=t}$ for the
conditional mean outcome of cell $\mathcal{C}$ in period
$t\in\{-K,\dots,0,1\}$. A cell is observed in every period, while the units
composing it and their covariates may differ across periods, which is what the
estimation of Section~\ref{sec:dml} has to accommodate.

\section{Identifying assumptions}\label{sec:ass}

\begin{assumption}[No anticipation and consistency]\label{A:noant}
$Y_t=Y_t(0,0)$ for $t\leq0$ and $Y_1=Y_1(D,G)$.
\end{assumption}

Assumption~\ref{A:noant} rules out that outcomes prior to the treatment respond to
the treatment of any unit, such that pre-treatment outcomes are realizations of
$Y_t(0,0)$ irrespective of a unit's eventual treatment and exposure. Under
interference the condition is stronger than its counterpart without spillovers,
since it also excludes anticipation of the treatment of others, but it concerns the
timing of the response alone and places no restriction on the interference
structure. It restricts outcomes, so units may learn of the treatment in advance as
long as their pre-treatment outcomes do not respond to it.

\begin{assumption}[Valid extrapolation]\label{A:ITS}
For a cell $\mathcal{C}$ and all $X$ in its support,
\begin{align}\label{forecast}
\Exp{Y_1(0,0)\mid X,\mathcal{C}}=\nu\!\left(X,\mathcal{C}\right),
\end{align}
where $\nu$ is a known functional of the pre-treatment outcomes of cell
$\mathcal{C}$ given $X$.
\end{assumption}

Assumption~\ref{A:ITS} states that the counterfactual conditional mean outcome in
the absence of any treatment can be predicted from the pre-treatment outcomes of the
same cell. For a polynomial trend fitted by least squares over the
pre-treatment periods, that is a regression of $\mu_t(X,\mathcal{C})$ on
$(1,t,\dots,t^P)$ whose fitted value at $t=1$ is taken as the forecast, $\nu$ is the
linear combination
\begin{align}\label{lintrend}
\nu(X,\mathcal{C})=\sum_{t=-K}^{0}w_t\,\mu_t(X,\mathcal{C}),
\end{align}
with projection weights $w_t$ that depend neither on $X$ nor on the cell. For $P=1$
and two pre-treatment periods this gives $\nu=2\mu_0-\mu_{-1}$, the leading case
below, equivalent to a constant mean change of $Y_t(0,0)$ across consecutive
periods, and $P$ may be chosen by cross-validation. More generally, $\nu$ may be any forecasting rule that is unbiased on average, see \citet{botosaru2023forecasted}, and further pre-treatment
periods permit placebo checks of the rule itself. With panel data $\nu$ may also be built from the unit's own outcome path, $\nu(X,\mathcal{C})=f(Y_0,\dots,Y_{-K},X,\mathcal{C})$ for a known $f$ satisfying \eqref{forecast}; in repeated cross sections weighted conditional means as in \eqref{lintrend} are the natural choice. The assumption restricts the evolution of $Y_t(0,0)$ within one cell, which is why interference leaves it intact. Conditioning on covariates makes it more plausible, allows trends
to differ with observed characteristics and permits the cell's covariate composition
to change across periods.

The next two assumptions transfer a counterfactual across cells and are needed only
for the decompositions of the total effect. Assumption~\ref{A:transfer} is an
effect homogeneity condition across treatment groups, Assumption~\ref{A:transfer2}
a selection-on-observables condition for exposure among the treated. The first
carries the spillover from the non-treated to the treated and the second the pure
direct effect from the non-exposed treated to all treated, which are the two routes
of Proposition~\ref{prop:direct}.

\begin{assumption}[Spillover homogeneity across treatment groups]\label{A:transfer}
For all $g$ and $X$ in the support among the treated,
\begin{align*}
\Exp{Y_1(0,g)-Y_1(0,0)\mid X,D=1,G=g}=\Exp{Y_1(0,g)-Y_1(0,0)\mid X,D=0,G=g},
\end{align*}
and $\Pr(D=0,G=g\mid X,T=1)>0$.
\end{assumption}

\begin{assumption}[Outcome homogeneity across exposure levels among the treated]
\label{A:transfer2}
For all $g$ and $X$ in the support among the treated,
\begin{align*}
\Exp{Y_1(1,0)\mid X,D=1,G=g}=\Exp{Y_1(1,0)\mid X,D=1,G=0},
\end{align*}
and $\Pr(G=0\mid X,D=1,T=1)>0$.
\end{assumption}

Assumption~\ref{A:transfer} states that treated and non-treated units sharing the
same exposure and covariates experience the same average spillover, and
Assumption~\ref{A:transfer2} that exposure is as good as randomly assigned among
the treated given covariates, as far as the outcome under non-exposure is
concerned. Both fail when the relevant quantity differs along unobserved
characteristics in which the compared cells differ, which is the case when treatment
and exposure are selected on such characteristics. Conditioning on a rich set of
covariates is the only remedy, which motivates the adjustment of
Section~\ref{sec:dml}. Their second parts require common support in exposure,
discussed in Section~\ref{sec:ident} and frequently the binding constraint.

\begin{assumption}[Conditional parallel trends]\label{A:CT}
For all $g$ and $X$ in the support among the treated,
\begin{align*}
\Exp{Y_1(0,0)-Y_0(0,0)\mid X,D=1}=\Exp{Y_1(0,0)-Y_0(0,0)\mid X,D=0,G=g},
\end{align*}
and $X$ has common support across the cells $\{D=1\}$ and $\{D=0,G=g\}$.
\end{assumption}

Assumption~\ref{A:CT} is a conditional parallel trends assumption, see e.g.
\citet{Abadie2005} and \citet{Lechner2010}, stated for the potential outcome in the
absence of any treatment. We impose it only as an optional
restriction, which overidentifies the spillover effect and generates a testable
implication.

\section{Identification}\label{sec:ident}

The total effect on the treated and the spillover on the non-treated follow from
trend stability within the respective cell alone. Neither requires a comparison
group, and the two are identified by the same argument applied to different cells.
The result is conditional on the covariates, and the aggregates follow by averaging
over the post-treatment covariate distribution of the cell considered.

\begin{prop}[Total and spillover effects]\label{prop:main}
Under Assumptions~\ref{A:noant} and \ref{A:ITS} for the cell $\mathcal{C}$,
\begin{align}\label{condid}
\tau_T^{1}(X)&=\mu_1\!\left(X,\cT\right)-\nu\!\left(X,\cT\right),\qquad
\cT=\{D=1\},\nonumber\\
\tau_I^{0}(X,g)&=\mu_1\!\left(X,\cN{g}{}\right)-\nu\!\left(X,\cN{g}{}\right),\qquad
\cN{g}{}=\{D=0,G=g\},
\end{align}
where $\cT$ and $\cN{g}{}$ denote the treated cell and the non-treated cell with
exposure $g$, and the aggregates follow by averaging over the post-treatment
covariate distribution of the respective cell.
\end{prop}

Identification thus requires nothing beyond the conditional mean outcomes of a
single cell: the post-treatment mean minus a prediction of what it would have been
had no unit been treated, formed from the cell's own pre-treatment means. No
comparison group and no common support condition across cells are needed, and
interference is left unrestricted beyond \eqref{expmap}. \citet{Cellietal2026}
establish both results without covariates; the conditional statement adds that
trends may differ across observed characteristics and that the covariate
composition of a cell may change across periods. Running through conditional means
rather than within-unit differences, it also applies to repeated cross sections,
requiring only that cell membership be observable in every period, which holds
whenever treatment and exposure are defined at the level of a region, village or
block.

The pre-treatment conditional means enter \eqref{condid} evaluated at the covariate
values of the post-treatment cell and must therefore be reweighted whenever the
covariate composition of the cell changes across periods. This is the generic case
in repeated cross sections and in unbalanced panels, and it is what the doubly
robust estimation of Section~\ref{sec:dml} delivers. Only in a balanced panel with
a unit-level covariate vector do the aggregates collapse, under \eqref{lintrend},
to group means of the within-unit contrast $S=Y_1-\sum_{t\leq0}w_tY_t$. For two
pre-treatment periods and a linear trend this contrast is the second difference
$Y_1-2Y_0+Y_{-1}$, and the aggregates are $\tau_T^{1}=\Exp{S\mid D=1}$ and
$\tau_I^{0}(g)=\Exp{S\mid D=0,G=g}$ without any covariate adjustment.

\begin{prop}[Decompositions of the total effect]\label{prop:direct}
Let Assumption~\ref{A:noant} and Assumption~\ref{A:ITS} hold for $\{D=1\}$.
\begin{itemize}
\item[(i)] If Assumption~\ref{A:ITS} holds for $\{D=0,G=g\}$ at each exposure level
occurring among the treated and Assumption~\ref{A:transfer} holds, then
\begin{align}\label{dirid}
\tau_I^{1}(0)=\textstyle\sum_{g\in\mathcal{G}}q_g\,
\Exp{\tau_I^{0}(X,g)\mid D=1,G=g,T=1},\qquad
\tau_D^{1}(G)=\tau_T^{1}-\tau_I^{1}(0),
\end{align}
where $q_g=\Pr(G=g\mid D=1,T=1)$.
\item[(ii)] If Assumption~\ref{A:ITS} holds for $\{D=1,G=0\}$ and
Assumption~\ref{A:transfer2} holds, then
\begin{align}\label{pureid}
\tau_D^{1}(X,0)&=\mu_1\!\left(X,\cTc{0}{}\right)-\nu\!\left(X,\cTc{0}{}\right),\qquad
\cTc{0}{}=\{D=1,G=0\},\nonumber\\
\tau_D^{1}(0)&=\Exp{\tau_D^{1}(X,0)\mid D=1,T=1},\qquad
\tau_I^{1}(1)=\tau_T^{1}-\tau_D^{1}(0).
\end{align}
\end{itemize}
\end{prop}

Each part delivers both components of a decomposition in \eqref{pure}, as the one
that is transferred determines the other as a residual from $\tau_T^{1}$: part~(i)
the interference effect on the treated absent their own treatment, the contrast of
$(0,G)$ with $(0,0)$, and part~(ii) the pure direct effect. Both transfer a
conditional effect from one cell to the covariate distribution of another, which is
where machine learning adjustment enters, and differ in the direction of the
transfer. By \eqref{de} and \eqref{pure}, neither decomposition requires a
restriction on the interaction between own treatment and exposure, since each holds
one argument fixed; the interaction is identified as $\tau_D^1(G)-\tau_D^1(0)$ when
both apply.

Common support in exposure is typically the binding constraint, and it binds
differently for the two parts: part (i) requires non-treated units at the exposure
levels occurring among the treated, part (ii) treated units that are not exposed at
all. Consider a conditional cash transfer program such as PROGRESA, where interference
operates within the village \citep{AngelucciDeGiorgi2009}. Eligible households in a
treated village receive the transfer, so a treated household linked only to
ineligible non-recipients may not exist and the common support condition in
Assumption~\ref{A:transfer2} fails. Assumption~\ref{A:transfer} remains plausible
there, since non-eligible households in treated villages are non-treated and as
exposed as the treated ones. Designs that do deliver common support randomize the
share of treated units across clusters.

Assumption~\ref{A:transfer2} extends part~(ii) from the non-exposed treated to all
treated. Without it, $\tau_D^{1}(X,0)$ in \eqref{pureid} and
$\tau_T^{1}(X)-\tau_D^{1}(X,0)$ still identify the pure direct and the interference
effect for the subpopulation $\{D=1,G=0\}$, which needs nothing beyond
Assumptions~\ref{A:noant} and \ref{A:ITS}. When common support fails, or when the
homogeneity conditions are implausible, the analysis is confined to the total effect
on the treated and the spillover effect on the non-treated, which rest on
Assumptions~\ref{A:noant} and \ref{A:ITS} alone. We regard these as the main
parameters and the decompositions as available only under favourable designs.
Finally, suppose the researcher is willing to impose Assumption~\ref{A:CT} in
addition, in which case the conditional spillover is overidentified.

\begin{prop}[Overidentification under parallel trends]\label{prop:alt}
Under Assumptions~\ref{A:noant}, \ref{A:ITS} for $\{D=1\}$ and \ref{A:CT},
\begin{align}\label{ctroute}
\tau_I^{0}(X,g)=\mu_1(X,\cN{g}{})-\mu_0(X,\cN{g}{})
-\left\{\nu(X,\cT)-\mu_0(X,\cT)\right\}.
\end{align}
If Assumption~\ref{A:ITS} holds for $\{D=0,G=g\}$ as well, expression
\eqref{ctroute} coincides with the expression of Proposition~\ref{prop:main} if and
only if $\nu(X,\cT)-\mu_0(X,\cT)=\nu(X,\cN{g}{})-\mu_0(X,\cN{g}{})$ almost surely.
\end{prop}

The two routes differ only in whose pre-treatment trend is used to extrapolate the
counterfactual of the non-treated cell, its own or that of the treated. They
therefore agree exactly when the conditional trends extrapolated from the
pre-treatment periods coincide across the two groups, which is what placebo tests in
conventional difference-in-differences analysis assess, and which by
Assumption~\ref{A:noant} involves only pre-treatment outcomes and can thus not be
contaminated by spillovers. We treat \eqref{ctroute} as a source of a testable restriction rather
than as a substitute for Proposition~\ref{prop:main}, since there is no reason to
expect the pre-treatment trend of the treated to approximate that of a non-treated
cell better than the cell's own.

\section{Doubly robust estimation by double machine learning}\label{sec:dml}

All parameters of Section~\ref{sec:ident} require averaging a conditional mean
identified in one cell over the covariate distribution of another: the
decompositions of Proposition~\ref{prop:direct} across treatment or exposure cells,
and the aggregates of Proposition~\ref{prop:main} across periods whenever a cell's
covariate composition changes. We therefore derive an expression that is doubly
robust \citep{Robins+94,RoRo95} and Neyman orthogonal \citep{Neyman1959}, which
permits machine learning estimation of the nuisance parameters. All objects of
interest are of the following form. Let $\mathcal{C}$ and
$\mathcal{A}$ be two disjoint cells defined by values of $(D,G,T)$, which we call
the donor and the target cell, let $V$ be an observed random variable and let
$R=\I\{\mathcal{A}\cup\mathcal{C}\}$ with $\I\{\cdot\}$ the indicator function. We
define within $R=1$ the probabilities $p=\Pr(\mathcal{A}\mid R=1)$ and
$\pi(X)=\Pr(\mathcal{A}\mid X,R=1)$ as well as the conditional mean
$m(X)=\Exp{V\mid X,\mathcal{C}}$. The parameter is
$\vartheta(\mathcal{A},\mathcal{C};V)=\Exp{m(X)\mid\mathcal{A}}$, a conditional
mean identified in the donor cell averaged over the covariate distribution of the
target cell. The covariates may include pre-treatment outcomes, which by
Assumption~\ref{A:noant} are unaffected by treatment and exposure and are typically
the most informative predictors here.

\begin{prop}[Doubly robust score]\label{prop:dr}
Let $\pi(X)$ be bounded away from zero and one on the support of $\mathcal{A}$ and
let $m$ be square integrable. Then $\vartheta(\mathcal{A},\mathcal{C};V)$ solves
$\Exp{\psi}=0$ for
\begin{align}\label{score}
\psi=\frac{\I\{\mathcal{A}\}}{p}\,m(X)
+\frac{\I\{\mathcal{C}\}\,\pi(X)}{\left(1-\pi(X)\right)p}
\left\{V-m(X)\right\}-\vartheta(\mathcal{A},\mathcal{C};V),
\end{align}
and $\psi$ is Neyman orthogonal with respect to $m$ and $\pi$.
\end{prop}

Each parameter of Section~\ref{sec:ident} is a linear combination of terms
\eqref{score} with $V=Y_t$ for a single period, as \eqref{score} is linear in $V$
and in $m$. Writing the exposure level as a superscript and the period as a subscript,
$\cT_t=\{D=1,T=t\}$, $\cTc{g}{t}=\{D=1,G=g,T=t\}$ and
$\cN{g}{t}=\{D=0,G=g,T=t\}$, the aggregate of
Proposition~\ref{prop:main} and the transfer in \eqref{dirid} are
\begin{align}\label{instances}
\tau_T^{1}&=\Exp{\mu_1(X,\cT)\mid \cT_1}-\Exp{\nu(X,\cT)\mid \cT_1}
=\Exp{Y_1\mid \cT_1}-\textstyle\sum_{t\leq0}w_t\vartheta(\cT_1,\cT_t;Y_t),\nonumber\\
\Exp{\tau_I^{0}(X,g)\mid \cTc{g}{1}}&=
\Exp{\mu_1(X,\cN{g}{})\mid \cTc{g}{1}}-\Exp{\nu(X,\cN{g}{})\mid \cTc{g}{1}}
\nonumber\\
&=\vartheta(\cTc{g}{1},\cN{g}{1};Y_1)
-\textstyle\sum_{t\leq0}w_t\vartheta(\cTc{g}{1},\cN{g}{t};Y_t).
\end{align}
The first equality in each case averages
\eqref{condid} over the covariates of the target cell by the law of iterated
expectations. The second inserts \eqref{lintrend} and uses that
$\mu_t(X,\mathcal{C})$ is the conditional mean $m(X)=\Exp{Y_t\mid X,\mathcal{C}_t}$
of the observed variable $V=Y_t$ in the donor cell $\mathcal{C}_t$, so that its
average over the covariates of the target cell is $\vartheta$. In the first line donor and target are the same cell in
different periods, so its own pre-treatment means are reweighted to its
post-treatment covariate distribution. At $t=1$ the two coincide and $\vartheta$ is
the cell mean of $Y_1$. In the term with $t=0$, for instance, $m(X)=\Exp{Y_0\mid X,D=1,T=0}$
is the mean outcome of the treated in the last pre-treatment period and $\vartheta$
is that conditional mean averaged over the covariates of the post-treatment
treated. In the second line the donor and the
target differ in treatment status at a common exposure level and period, so the
conditional spillover is carried from the non-treated to the treated. In a balanced
panel with a unit-level covariate vector the combination can be formed within units,
so each line reduces to a single term with $V=S$.
The remaining parameters are of the same form. Appendix~\ref{app:inst} gives the
split of $\tau_D^{1}(0)$ by exposure used in \eqref{pureid} and the cells behind the
two test statistics of Section~\ref{sec:test}, each of which is again a linear
combination of terms \eqref{score} with a donor and a target cell defined by values
of $(D,G,T)$.

Estimation proceeds by cross-fitting as in \citet{Chetal2018}. We split the sample
into $L$ folds and estimate $m$ by regressing $V$ on $X$ in the donor cell, and
$\pi$ within $R=1$, in all folds but one. Any machine learner satisfying the rate
condition of Assumption~\ref{A:reg} may be used, such as penalized regression
\citep{tihonov1963solution,Bellonietal2014}, random forests \citep{WagerAthey2018}, neural networks
\citep{FarrellLiangMisra2018} or an ensemble \citep{VanderLaanHubbard2007}. We then
evaluate \eqref{score} in the held-out fold and average over folds. The conditional
means entering \eqref{ctroute} are estimated in the same way. Cross-fitting removes the
own-observation bias of the nuisance estimates, so that the rate condition of
Assumption~\ref{A:reg}(d) is all the machine learner must satisfy.

\begin{assumption}[Regularity]\label{A:reg}
(a) Units are sampled independently and the outcomes have finite fourth moments.
(b) $\pi(X)$ and $p$ are bounded away from zero and one on the relevant support. (c) The conditional variance of $V$ given $X$ is
bounded away from zero. (d) The nuisance estimators are consistent and satisfy
$\|\hat\pi-\pi\|_2\cdot\|\hat m-m\|_2=o_p(n^{-1/2})$.
\end{assumption}

Interference induces dependence across units in the outcomes, such that
Assumption~\ref{A:reg}(a) refers to the sampling of units and their
characteristics, with exposure a fixed function of the realized assignment, as in
the assignment-conditional perspective of \citet{Savjeetal2021}. Under spatial or
network dependence, the variance is clustered at the level at which the exposure
mapping operates. In the application of Section~\ref{sec:app} this is the branch of
the patent classification, within which a subclass can reach another.

\begin{theorem}[Asymptotic normality]\label{thm:1}
Under Assumptions~\ref{A:noant}, \ref{A:ITS} and \ref{A:reg}, the cross-fitted
estimators of $\tau_T^{1}$, $\tau_I^{0}(g)$ and
$\vartheta(\mathcal{A},\mathcal{C};V)$ are
$\sqrt{n}$-consistent and jointly asymptotically normal, with a covariance matrix
given by that of the corresponding scores and consistently estimated by its
empirical analogue. The same holds for $\tau_D^{1}(G)$ under
Assumption~\ref{A:transfer} and for $\tau_D^{1}(0)$ and $\tau_I^{1}(1)$ under
Assumption~\ref{A:transfer2}, by the delta method applied to \eqref{dirid} and
\eqref{puresplit}.
\end{theorem}

Theorem~\ref{thm:1} takes the weights $w_t$ in \eqref{lintrend} as known, which holds for a polynomial trend of given order $P$. When $P$ is selected by cross-validation, the candidate orders are bounded by the number of pre-treatment periods and thus form a finite set. The asymptotic distribution is unaffected if the population cross-validation criterion has a unique minimizer $P^*$, such that its value at any other order exceeds that at $P^*$ by a positive margin. Any consistent estimator of the criterion then selects $P^*$ with probability approaching one. On this event, the estimator coincides with the one based on the known order $P^*$, and the two therefore share the limiting distribution in Theorem~\ref{thm:1}. 

The weights may also depend on an estimated parameter. An example is the damped trend $\nu(X,\mathcal{C})=\mu_0(X,\mathcal{C})+\kappa_{\mathcal{C}}\{\mu_0(X,\mathcal{C})-\mu_{-1}(X,\mathcal{C})\}$, with weights $w_0=1+\kappa_{\mathcal{C}}$ and $w_{-1}=-\kappa_{\mathcal{C}}$. It nests the level rule for $\kappa_{\mathcal{C}}=0$ and the linear rule for $\kappa_{\mathcal{C}}=1$. The parameter $\kappa_{\mathcal{C}}$ is the autoregressive coefficient of the first differences of the mean outcomes of cell $\mathcal{C}$ across the pre-treatment periods, which requires at least three such periods. If $\hat\kappa_{\mathcal{C}}$ is $\sqrt{n}$-consistent, all estimators involving $\nu(X,\mathcal{C})$ remain $\sqrt{n}$-consistent and asymptotically normal. Each such parameter contains the term $\sum_{t\le 0} w_t(\kappa_{\mathcal{C}})\,\vartheta(\mathcal{A},\mathcal{C}_t;Y_t)$, where $\mathcal{A}$ denotes the target cell and $\mathcal{C}_t$ the cell $\mathcal{C}$ in period $t$. Its influence function therefore gains the term $-\big\{\sum_{t\le 0}\vartheta(\mathcal{A},\mathcal{C}_t;Y_t)\,\partial w_t(\kappa_{\mathcal{C}})/\partial\kappa_{\mathcal{C}}\big\}\,\phi_{\kappa_{\mathcal{C}}}$, where $\phi_{\kappa_{\mathcal{C}}}$ denotes the influence function of $\hat\kappa_{\mathcal{C}}$. This term is of the same order as the score and enters the asymptotic variance. Standard errors are therefore obtained from the sum of the influence functions.

\section{Testable restrictions}\label{sec:test}

Since the spillover is zero for units classified as non-exposed, the framework
implies the restriction
\begin{align}\label{R1}
H_0^{(1)}:\quad \tau_I^{0}(X,0)=0\quad\text{almost surely.}
\end{align}
A rejection indicates either that Assumption~\ref{A:ITS} fails for that cell or
that \eqref{expmap} is not a sufficient summary of the interference structure, such
that units classified as non-exposed are in fact reached by the treatment. The
restriction thus assesses the design jointly with the exposure mapping, and requires
the mapping to claim that some units are non-exposed but not that the claim be
correct. We state it conditional on $X$, because a misclassification concentrated in
a region of the covariate space may be small on average.

A second restriction follows from Proposition~\ref{prop:alt}. If
Assumption~\ref{A:CT} is imposed in addition to trend stability in both
subpopulations, the conditional spillover is overidentified, and the two routes
agree if and only if
\begin{align}\label{R2}
H_0^{(2)}:\ \nu(X,\cN{g}{})-\mu_0(X,\cN{g}{})=\nu(X,\cT)-\mu_0(X,\cT)\ \text{ a.s.,}
\end{align}
the equality across groups of the conditional trends extrapolated from the
pre-treatment periods, which for two such periods and a linear trend is the equality
of the conditional pre-treatment first differences. This is what placebo tests
assess in conventional difference-in-differences analysis.

We aggregate the conditional restriction into a pair of global statistics, building on
 \citet{Apfel2023learning}, see \cite{ArmendarizHuber2026} for a related application. Let
$\delta(X)$ denote the conditional quantity restricted to zero under the respective
null and define
\begin{align}\label{globaltheta}
\vartheta^{\pm}=\Exp{\delta(X)^2\pm\delta(X)\mid D=1},
\end{align}
both of which are zero under the null. The quadratic term makes the test consistent against all alternatives. The linear term ensures a non-degenerate limiting distribution under the null \citep{Apfel2023learning}. The pair is used because a single statistic vanishes whenever the quadratic and linear terms offset each other. The orthogonal score is
\begin{align}\label{globalscore}
\psi^{\pm}=\frac{D}{p}\left\{\delta(X)^2\pm\delta(X)\right\}
+\left\{2\delta(X)\pm1\right\}a(W)-\vartheta^{\pm},
\end{align}
where $a(W)$ is the second term in \eqref{score} with $m=\delta$, the target cell
being $\mathcal{A}=\{D=1\}$ in both cases and the donor cell
$\mathcal{C}=\{D=0,G=0\}$ for \eqref{R1} and the cells listed in
Appendix~\ref{app:inst} for \eqref{R2}, and $p$ is the share of treated units in
$R=1$. Under Assumption~\ref{A:reg}, the additional conditions
$\|\hat\delta-\delta\|_2=o_p(n^{-1/4})$ and $\|\hat\pi-\pi\|_2=o_p(n^{-1/4})$,
and the moment bounds on the nuisance estimators in
\citet[Assumption~6]{Apfel2023learning}, the cross-fitted estimates of
$\vartheta^{+}$ and $\vartheta^{-}$ are jointly asymptotically normal and mirror
images of each other under the null. Then $\max\{t^{+},t^{-}\}$ converges to $|Z|$ with $Z\sim N(0,1)$ and
the test rejects at level $\alpha$ if $\max\{t^{+},t^{-}\}>z_{1-\alpha/2}$. Joint
testing across exposure cells stacks the statistics and applies
\citet{romano2005exact}. Joint validity of the assumptions implies
the restriction; the converse requires excluding that two non-zero biases coincide,
a faithfulness condition in the sense of \citet{Pearl00} invoked for the same
purpose for instance in \citet{huber2023testing} and \citet{HuberOess2025}. A zero
$\tau_I^{0}(X,g)$ where exposure is believed positive is consistent with no
spillover and with a spillover offsetting a failure of trend stability, which is why
estimates are reported across all exposure cells.

\section{Simulation study}\label{sec:sim}

In this section, we consider a simulation study based on repeated cross sections.
We draw $n$ units in clusters of five, $Q=20$
independent standard normal covariates $X$ fixed at the unit level and an
unobservable $U=0.7X_1-0.5X_2+\epsilon$. A cluster is activated if
$0.6\bar{X}_1+\eta>0$, with $\bar{X}_1$ its mean of $X_1$ and $\eta$ a
cluster-level shock, and a unit in an activated cluster is treated if
$0.5X_1+0.3X_3-0.2X_4+U+\zeta>0$ with $\zeta$ logistic, so selection depends on
unobservables. Exposure is binary, $G=\I\{$at least one other unit in the cluster
is treated$\}$, so non-treated units in activated clusters are exposed and exposure
has common support across treatment groups. Potential outcomes in the absence of any treatment are
$Y_t(0,0)=X'\beta+\lambda_tU+\alpha+\varepsilon_t$ for $t\in\{-1,0,1\}$, that is
$K=1$, with $\beta_q=0.5/q$ and $(\lambda_{-1},\lambda_0,\lambda_1)=(0,a,2a+c)$. The
observed post-treatment outcome adds a direct effect of one, a spillover
$(1+0.5X_5)G$ and an interaction $0.5\,DG$, so Assumption~\ref{A:transfer} holds
while own treatment and exposure interact. Each unit enters
period $t$ with probability $[1+\exp\{-(0.8tX_6-0.4t)\}]^{-1}$ and only the
selected rows are observed, so the covariate composition of every cell differs
across periods and no within-unit difference can be formed.

As the groups differ in their composition in $U$ given $X$, the conditional
pre-treatment trend of a cell is $a\Exp{U\mid X,\mathcal{C}}$, so
Assumption~\ref{A:CT} and hence \eqref{R2} hold if and only if $a=0$. The second
difference of the loading is $c$, so Assumption~\ref{A:ITS} holds if and only if
$c=0$. We report all four combinations of $a\in\{0,0.5\}$ and $c\in\{0,0.4\}$.
With two pre-treatment periods and a linear trend, \eqref{lintrend} is
$\nu(X,\mathcal{C})=2\mu_0(X,\mathcal{C})-\mu_{-1}(X,\mathcal{C})$: each cell is
extrapolated from its own two conditional means
$\mu_t(X,\mathcal{C})=\Exp{Y\mid X,\mathcal{C},T=t}$, each a cross-fitted ridge
regression \citep{tihonov1963solution} on the rows of that cell and period. The parameters estimated are
$\tau_T^{1}$ and $\tau_I^{0}(1)$ as in \eqref{instances} and the direct effect
$\tau_D^{1}(G)=\tau_T^{1}-q\Exp{\tau_I^{0}(X,1)\mid\cTc{1}{1}}$ of
Proposition~\ref{prop:direct}(i), where \eqref{dirid} has no term at $g=0$, with
population values $2.42$, $1.00$ and $1.47$. Estimation follows Section~\ref{sec:dml}.
\ifnum\arxiv=0
The learners, the trimming and the
function $\delta(X)$ that \eqref{globaltheta} aggregates into the test statistic are
stated in Appendix~\ref{app:sim}.
\fi

\ifnum\arxiv=1
The nuisances of the design are estimated as follows.
Every $m$ is a cross-fitted ridge regression on the
rows of the donor cell and every $\pi$ a cross-fitted $\ell_2$-penalized logistic
regression on the union of
the two cells, with three folds and propensity scores trimmed at $0.02$ and $0.98$
as in \citet{CrumpImbensMitnik09} and \citet{HuLeWu10}; each $\vartheta$ is the
sample mean of \eqref{score}, the aggregates are the linear combinations in
\eqref{instances} with standard errors from the corresponding influence functions,
and $\tau_D^{1}(G)$ follows by the delta method. For the test of \eqref{R2},
$\delta(X)$ is the difference between the extrapolated conditional trends of
$\{D=0,G=1\}$ and of the treated, evaluated on the post-treatment treated.
\ifnum\arxiv=0
We run 300 replications for $n\in\{4000,16000\}$.
\fi

\begin{table}[t]
\centering\scriptsize\setlength{\tabcolsep}{3pt}
\caption{Monte Carlo results, 300 replications}\label{tab:sim}
\begin{tabular}{llccccccccccccc}
\toprule
 & & \multicolumn{4}{c}{$\tau_T^{1}$} & \multicolumn{4}{c}{$\tau_I^{0}(1)$} & \multicolumn{4}{c}{$\tau_D^{1}(G)$} & Test \\
\cmidrule(lr){3-6}\cmidrule(lr){7-10}\cmidrule(lr){11-14}\cmidrule(lr){15-15}
$(a,c)$ & $n$ & bias & sd & rmse & avse & bias & sd & rmse & avse & bias & sd & rmse & avse & \eqref{R2} \\
\midrule
(0,0) & 4000 & -0.001 & 0.156 & 0.156 & 0.177 & -0.012 & 0.177 & 0.177 & 0.188 & 0.017 & 0.301 & 0.301 & 0.322 & 0.000 \\
 & 16000 & -0.007 & 0.073 & 0.074 & 0.079 & -0.002 & 0.086 & 0.086 & 0.084 & -0.001 & 0.144 & 0.143 & 0.149 & 0.000 \\
(0.5,0) & 4000 & -0.007 & 0.164 & 0.164 & 0.186 & -0.014 & 0.182 & 0.182 & 0.197 & 0.010 & 0.314 & 0.314 & 0.343 & 0.037 \\
 & 16000 & -0.008 & 0.076 & 0.076 & 0.084 & -0.002 & 0.093 & 0.092 & 0.089 & 0.002 & 0.150 & 0.150 & 0.157 & 0.717 \\
(0,0.4) & 4000 & 0.292 & 0.157 & 0.331 & 0.178 & -0.274 & 0.177 & 0.326 & 0.189 & 0.353 & 0.306 & 0.467 & 0.324 & 0.000 \\
 & 16000 & 0.286 & 0.075 & 0.296 & 0.080 & -0.266 & 0.088 & 0.280 & 0.085 & 0.337 & 0.144 & 0.366 & 0.150 & 0.000 \\
(0.5,0.4) & 4000 & 0.286 & 0.167 & 0.331 & 0.190 & -0.276 & 0.185 & 0.332 & 0.201 & 0.346 & 0.325 & 0.475 & 0.351 & 0.037 \\
 & 16000 & 0.285 & 0.078 & 0.296 & 0.086 & -0.266 & 0.095 & 0.282 & 0.091 & 0.340 & 0.153 & 0.373 & 0.161 & 0.717 \\
\bottomrule
\end{tabular}
\begin{minipage}{0.98\textwidth}\vspace{3pt}\scriptsize
\textit{Notes:} `bias', `sd' and `rmse' are the Monte Carlo bias, standard deviation and root mean squared error
of the estimator relative to the population effect, `avse' is the average estimated standard error, and the last
column gives the rejection rate of the test of \eqref{R2} at the five percent level, which holds if and only if
$a=0$. Assumption~\ref{A:ITS} holds if and only if $c=0$. The number of observed unit-period rows is about
$1.5n$. Nuisances are estimated by cross-fitted ridge and penalized logistic
regression, $Q=20$.
\end{minipage}
\end{table}
\fi

Table~\ref{tab:sim} reports the results of 300 replications for
$n\in\{4000,16000\}$. When $c=0$ the three estimators are close to unbiased and
their root mean squared error halves when the number of units quadruples, as implied by the $\sqrt{n}$-rate
of Theorem~\ref{thm:1}. The average estimated standard error is within a tenth of
the Monte Carlo standard deviation, so the influence functions of
Section~\ref{sec:dml} deliver reliable inference in a design in which every
aggregate requires reweighting across periods. When $c=0.4$ all three estimators are biased by about $0.3$, with
opposite signs for the total and the spillover effect, because the treated and the exposed
non-treated differ in their composition in $U$, which drives the missed curvature.
The overidentification test behaves as intended and responds to $a$ alone: it never
rejects \eqref{R2} in the $600$ replications with $a=0$, so it is conservative under
the null, and rejects in $3.7$ percent at $n=4000$ and $71.7$ percent at $n=16000$
when $a=0.5$. A violation of Assumption~\ref{A:ITS} through $c$ leaves the
rejection rate unchanged, since the statistic uses pre-treatment outcomes alone.

\section{Empirical application}\label{sec:app}

We revisit the compulsory licensing of enemy-owned patents under the Trading with
the Enemy Act with the data of \citet{MoserVoena12}: the $7{,}248$ chemical
subclasses of the United States patent classification, with domestic patent counts
as the outcome and licensing in $336$ subclasses in 1919. The trend is fitted on the
$15$ pre-treatment periods 1904 to 1918, with 1899 to 1903 supplying the
covariates. \citet{MoserVoena12} regress
patents on treatment interacted with the post-1918 indicator, with subclass and year
fixed effects and a foreign-patent control, reporting $0.151$ to $0.255$ additional
patents per year. Their comparison group is every subclass without a licensed patent
and so contains those closest to the treated, where a spillover is most likely,
while identification rests on that group being unaffected. We define exposure from
the classification itself: a subclass is exposed if its branch, a main class
combined with the first two digits of the subclass code, contains a licensed
subclass other than itself. On this definition $1{,}594$ non-treated subclasses are
exposed and $5{,}318$ are not, and a share $q=\Pr(G=1\mid D=1,T=1)=0.72$ of the
treated are exposed as well.

\ifnum\arxiv=1
\ifnum\arxiv=1
The panel records for each subclass the number of patents granted in each year,
separately for domestic and for foreign inventors, and an indicator of whether the
subclass contains a patent licensed under the Trading with the Enemy Act. The mean
of the outcome over 1919 to 1939 is $0.62$ patents per subclass and year.
\else
The data are the chemical patent panel of \citet{MoserVoena12}, which records for
each of the $7{,}248$ subclasses of the United States patent classification in
chemicals the number of patents granted in each year, separately for domestic and
for foreign inventors, and an indicator of whether the subclass contains a patent
licensed under the Trading with the Enemy Act. The outcome is the domestic count,
whose mean over 1919 to 1939 is $0.62$ patents per subclass and year. Licensing took
place in 1919 and affected $336$ subclasses. The trend is fitted on the $15$
pre-treatment periods 1904 to 1918, and the five years 1899 to 1903 preceding that
window supply the covariates.

\fi
\ifnum\arxiv=0
The exposure mapping uses the hierarchy of the classification. A branch is a main
class combined with the first two digits of the subclass code, of which there are
$824$ with a median of nine subclasses. A non-treated subclass has $G=1$ if its
branch contains a licensed subclass, and a treated subclass has $G=1$ if its branch
contains some other licensed subclass, which holds for $72$ percent of them, so that
$\cTc{0}{}$ is not empty and the common support required by
Proposition~\ref{prop:direct} is available.
\else
The $824$ branches hold a median of nine subclasses each, and $28$ percent of the
treated subclasses are not exposed, so that $\cTc{0}{}$ is not empty and the common
support required by Proposition~\ref{prop:direct} is available.
\fi
All $19$ main classes contain licensed
subclasses, so the main class is too coarse a notion of proximity to leave a
non-exposed cell, while the full subclass code is the unit itself; the branch is the
coarsest level in between. The $23$ covariates are the logged average number of own
and of foreign patents over 1899 to 1901 and over 1902 to 1903, the logged branch
size and $18$ main-class indicators. They are pre-treatment outcomes and
characteristics, which Assumption~\ref{A:noant} makes admissible as covariates, and
they lie before the fitting window of every specification reported here. For the
transfer from the exposed non-treated to the exposed treated the estimated
propensity score ranges from $0.02$ to $0.82$ with a cross-fitted McFadden
pseudo-$R^2$ of $0.10$, and for the transfer from the non-exposed cell to the
$94$ non-exposed treated from $0.02$ to $0.27$ with $0.02$, so the trimming of
Assumption~\ref{A:reg}(b) binds on few observations.
\fi

We estimate $\nu$ by \eqref{lintrend} as a linear trend in calendar time, evaluated
at the mean of 1931 to 1939\ifnum\arxiv=0 ; cross-validation within the
pre-treatment era selects the linear rule over the alternative orders\fi.
\citet{MoserVoena12} find a lag of six
to nine years and attribute it to the time domestic firms needed to acquire the
knowledge and skills to work the licensed inventions, and to the two to three years
between application and grant. We do not treat the
non-exposed subclasses as a cell that licensing leaves untouched, since the
classification is only a proxy for technological proximity, so \eqref{dirid} is
evaluated at both exposure levels. The data form a balanced panel, so by
Proposition~\ref{prop:main} the first three effects are means of the within-unit
contrast $S=Y_1-\sum_{t\leq0}w_tY_t$ in their own cell. The decomposition of
Proposition~\ref{prop:direct}(i) transfers each conditional spillover to the treated
at the same exposure level and is therefore a doubly robust term of
Section~\ref{sec:dml},
\begin{align}\label{appspec}
\tau_T^{1}&=\Exp{S\mid\cT_1},\qquad
\tau_I^{0}(1)=\Exp{S\mid\cN{1}{1}},\qquad
\tau_I^{0}(0)=\Exp{S\mid\cN{0}{1}},\nonumber\\
\tau_I^{1}(0)&=q\,\vartheta(\cTc{1}{1},\cN{1}{1};S)
+(1-q)\,\vartheta(\cTc{0}{1},\cN{0}{1};S),\qquad
\tau_D^{1}(G)=\tau_T^{1}-\tau_I^{1}(0),
\end{align}
with $m(X)=\Exp{S\mid X,\mathcal{C}}$ in the donor cell $\mathcal{C}$. A conditional
mean $m$ and a propensity score $\pi$ are therefore needed only in the two
transfers, where the donor and the target cell differ.
\ifnum\arxiv=0
Both are estimated by
cross-fitted ridge regression \citep{tihonov1963solution} and a cross-fitted
$\ell_2$-penalized logistic regression in the number of own and of foreign patents
over 1899 to 1901 and over 1902 to 1903, the size of the branch and indicators for
the $19$ main classes.
\fi
Standard errors are clustered at the branch level rather than
at the subclass, the unit of observation, because the exposure mapping operates
within branches, as discussed after Assumption~\ref{A:reg}.

\ifnum\arxiv=1
Figure~\ref{fig:extrap} plots the mean outcome of each cell together with the linear
rule fitted on 1904 to 1918 and extended through 1939. The treated and the exposed
non-treated lie above their own extrapolated trends in the 1930s, by $0.80$ and
$0.35$ patents per subclass and year. The non-exposed cell lies $0.03$ above its
own trend\ifnum\arxiv=0 , and evaluating that cell at the covariate distribution of
the non-exposed treated raises the gap to $0.286$\fi.

\begin{figure}
\centering
\includegraphics[width=\textwidth]{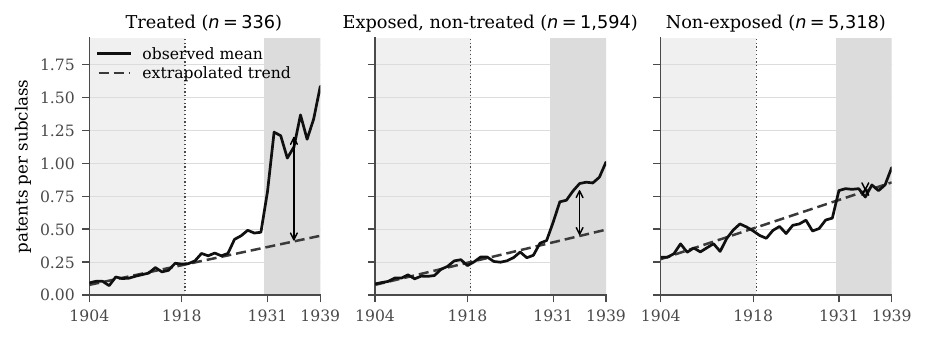}
\caption{Mean domestic patents per subclass and year by cell, with the linear rule
fitted on 1904--1918 (light band) and extrapolated through 1939. The arrow is the
gap over the post-treatment window 1931--1939 (dark band); the dotted line marks the
licensing year 1919.}\label{fig:extrap}
\end{figure}

The order $P$ of the polynomial in \eqref{lintrend} is selected by cross-validation
inside the pre-treatment era. For each of the origins $t_0=1907,\dots,1910$ the
polynomial of order $P$ is fitted on the years from 1904 to $t_0$ and used to
predict the mean outcome over the years $t_0+4$ to $t_0+8$, which is then compared
with the realised mean in each cell. The validation years always follow the fitting
years and no outcome after 1918 is used, so the exercise mimics the forecast the
application performs, over fitting windows of four to seven years and a horizon of
four to eight years. These are the origins at which the fitting window holds at
least four years, so that every order is identified at every origin. The polynomial
is fitted in centred and scaled time, since on raw calendar years the basis
$(1,t,\dots,t^P)$ has a condition number above $10^{11}$ already at $P=2$ and a
least squares routine then returns the weights of the linear rule.

Table~\ref{tab:cv} reports the results. The level rule $P=0$ leaves a mean
prediction error of $0.125$ in the non-exposed cell and $0.086$ among the treated,
which the linear rule reduces to $-0.028$ and $0.058$. The quadratic rule attains a
comparable bias but is five to seven times as dispersed across subclasses, and the cubic rule
is worse on both counts. We therefore use $P=1$ throughout.

\begin{table}
\centering\small
\caption{Cross-validation of the order $P$ inside the pre-treatment era}\label{tab:cv}
\begin{tabular}{lcccccc}
\toprule
& \multicolumn{3}{c}{mean prediction error} & \multicolumn{3}{c}{root mean squared error} \\
\cmidrule(lr){2-4}\cmidrule(lr){5-7}
$P$ & $\cT$ & $\cN{1}{}$ & $\cN{0}{}$ & $\cT$ & $\cN{1}{}$ & $\cN{0}{}$ \\
\midrule
0 & 0.086 & 0.092 & 0.125 & 0.47 & 1.10 & 0.78 \\
1 & 0.058 & -0.008 & -0.028 & 1.30 & 1.33 & 1.69 \\
2 & 0.065 & -0.029 & -0.046 & 6.55 & 7.01 & 11.07 \\
3 & -0.771 & 0.131 & 1.932 & 60.11 & 65.82 & 111.15 \\
\bottomrule
\end{tabular}
\begin{minipage}{0.84\textwidth}\vspace{3pt}\small
\textit{Notes:} $\cT$ denotes the treated subclasses, $\cN{1}{}$ the non-treated
subclasses that are exposed and $\cN{0}{}$ those that are not. The polynomial of
order $P$ is fitted on $[1904,t_0]$ and used to predict the mean outcome over
$[t_0+4,t_0+8]$, for the four origins $t_0=1907,\dots,1910$. The left block reports
the mean prediction error within each cell, the right block its root mean square
across subclasses, both averaged over origins. $P=0$ is the level rule $\nu=\mu_0$.
No outcome after 1918 is used.
\end{minipage}
\end{table}

The estimators are those of \eqref{appspec} and the nuisance functions are estimated
as in the simulation, every $m$ by a cross-fitted ridge regression on the
rows of the donor cell and every $\pi$ by a cross-fitted $\ell_2$-penalized logistic
regression on the union of the two cells, with the same trimming at $0.02$ and
$0.98$ and with five folds in place of three. Both transfers depend on the
assignment of subclasses to folds, which matters here because only $94$ treated
subclasses are non-exposed. Each is therefore reported as the median over nine fold
draws, and its standard error as the median of
$(\hat\sigma_s^2+(\hat\vartheta_s-\hat\vartheta)^2)^{1/2}$ across those draws.
This is the aggregation of \citet{Chetal2018}.
\ifnum\arxiv=0
The total effect on the treated and the
two spillover effects on the non-treated are means of $S$ in their own cell, without
covariate adjustment, since the donor and the target cell coincide and the score
\eqref{score} reduces to $S$ itself. The interference effect on the treated is the
sum $q\,\vartheta(\cTc{1}{1},\cN{1}{1};S)+(1-q)\,\vartheta(\cTc{0}{1},\cN{0}{1};S)$,
in which the conditional mean $m(X)=\Exp{S\mid X,\mathcal{C}}$ and the propensity
score enter because donor and target differ, and the direct effect is the residual
$\tau_T^{1}-\tau_I^{1}(0)$.
\fi

Table~\ref{tab:appdetail} reports the estimates of the effects and the placebo
estimates. Licensing raised domestic patenting in a treated subclass by $0.799$
$(0.105)$ patents per year, which is significant at the one percent level. In an
exposed non-treated subclass, patenting rose by $0.354$ $(0.138)$, which is
significant at the five percent level. Both effects are sizeable relative to the
mean of $0.62$ patents per subclass and year over 1919 to 1939. The non-exposed
subclasses gain $0.032$ $(0.048)$ on average, which is statistically
insignificant. Using all non-treated subclasses as the comparison group, the
estimated effect of licensing on the treated subclasses amounts to $0.693$
$(0.107)$. By construction, this falls short of the total effect by the average
spillover among the non-treated, which amounts to $0.107$ $(0.049)$ and is
significant at the five percent level. The decomposition attributes $0.418$ $(0.120)$ of the total effect on the treated to the direct channel and $0.382$ $(0.094)$ to exposure, both significant at the one percent level. The direct channel thus accounts for slightly more than half of the total effect. The second component exceeds $q\,\tau_I^{0}(1)=0.255$ because
each transfer evaluates the spillover of a non-treated cell at the covariates of
the treated with the same exposure level. At these covariates, the exposed cell
contributes $0.419$ $(0.128)$ rather than its unweighted mean of $0.354$, and the
non-exposed cell contributes $\vartheta(\cTc{0}{1},\cN{0}{1};S)=0.286$ $(0.064)$
rather than $0.032$. The spillover in the non-exposed cell is thus close to zero on
average but positive at the covariate values of the non-exposed treated, so both
non-treated groups are contaminated.

\ifnum\arxiv=1
The placebo test in Table~\ref{tab:appdetail} applies the same estimators entirely
within the pre-treatment era, fitting the rule on the thirteen years 1904 to 1916
and extrapolating to 1917 and 1918. The total
effect on the treated is $0.030$ $(0.028)$ and the direct effect $-0.001$ $(0.032)$,
against $0.80$ and $0.42$ in the main specification, and the two spillover effects
on the non-treated are $0.011$ $(0.016)$ and $0.008$ $(0.013)$. Only the
interference effect on the treated, $0.031$ $(0.012)$, differs from zero, and it is
a twelfth of the $0.382$ of the main specification.

\begin{table}
\centering\small\setlength{\tabcolsep}{4pt}
\caption{Compulsory licensing: estimates and placebo test}\label{tab:appdetail}
\begin{tabular}{llccccc}
\toprule
pre & post & $\tau_T^{1}$ & $\tau_D^{1}(G)$ & $\tau_I^{1}(0)$
 & $\tau_I^{0}(1)$ & $\tau_I^{0}(0)$ \\
\midrule
1904--1918 & 1931--1939 & 0.799$^{***}$ & 0.418$^{***}$ & 0.382$^{***}$ & 0.354$^{**}$ & 0.032 \\
 &  & (0.105) & (0.120) & (0.094) & (0.138) & (0.048) \\
\midrule
\multicolumn{7}{l}{\textit{placebo: extrapolation within the pre-treatment era}}\\
1904--1916 & 1917--1918 & 0.030 & -0.001 & 0.031$^{***}$ & 0.011 & 0.008 \\
 &  & (0.028) & (0.032) & (0.012) & (0.016) & (0.013) \\
\bottomrule
\end{tabular}
\begin{minipage}{0.95\textwidth}\vspace{3pt}\small
\textit{Notes:} both rows use the linear rule $P=1$ and the covariates described
above, with branch-clustered standard errors in parentheses and the two transfers
aggregated over nine fold draws. $\tau_I^{0}(1)$ and
$\tau_I^{0}(0)$ are the mean contrasts in the exposed and in the non-exposed
non-treated cell, while the decomposition reported under $\tau_D^{1}(G)$ and
$\tau_I^{1}(0)$ transfers each of them to the treated at the same exposure level, as
in \eqref{appspec}. Stars denote significance at the ten ($^{*}$), five ($^{**}$) and
one ($^{***}$) percent level, based on the normal approximation to the ratio of the
estimate to its standard error.
\end{minipage}
\end{table}
\fi

\section{Conclusion}\label{sec:conclusion}

\ifnum\arxiv=1
Under interference, comparison-group methods identify only the difference between
the total effect on the treated and the spillover effect on the non-treated. We have
shown that extrapolating the counterfactual conditional mean within a group
identifies both effects separately, in repeated cross sections as in panels and
without uncontaminated comparison units. The argument rests on no anticipation and
on a valid extrapolation of the pre-treatment trend of the cell considered.
Decomposing the total effect into a direct and an interference component is more
demanding. It requires effect homogeneity conditional on observables across treated
and non-treated units at the same exposure level, as well as common support in
exposure, which is why we regard the total effect on the treated and the spillover
on the non-treated as the main parameters of the framework.

The framework delivers two testable restrictions. The spillover must be zero for the
units the exposure mapping classifies as non-exposed, which assesses the identifying
assumptions jointly with the measurement of exposure. If conditional parallel trends
are imposed in addition, the conditional spillover is overidentified, and the two
routes agree if and only if the extrapolated pre-treatment trends coincide across
groups. We derived doubly robust and Neyman-orthogonal expressions for all
parameters, which permit estimating the conditional means and the propensity scores
by machine learning, and established root-$n$ consistency and asymptotic normality.
The simulation study confirms this behaviour and shows the overidentification test
to have power against violations of conditional parallel trends.

In the application to the compulsory licensing of enemy-owned patents in the United
States, licensing raised domestic patenting in a licensed subclass by $0.80$ patents
per year and in a technologically close one by $0.35$. Against all non-treated
subclasses the treated gain $0.69$, which understates the total effect by the
average spillover in that comparison group. Two limitations remain. The
decomposition rests on an effect homogeneity condition that the data cannot test,
and the extrapolation over a horizon of thirteen to twenty-one years is only as good
as the linear trend it assumes, which the placebo test checks at shorter horizons
alone.
\else
Under interference, comparison-group methods identify only the difference between
the total effect on the treated and the spillover effect on the non-treated.
Extrapolating the counterfactual conditional mean within a group identifies both
separately, in repeated cross sections as in panels and without uncontaminated
comparison units. The decomposition into the two components further requires effect
homogeneity conditional on observables and common support in exposure.
\fi

\clearpage

\bibliography{lit_spill}

\paragraph{Use of AI tools.} Generative AI was used to improve the language of the
manuscript and to assist with the code for the simulation and the application.

\appendix
\numberwithin{equation}{section}
\section{Instantiations of the doubly robust score}\label{app:inst}

Two further instances complete the list. For \eqref{pureid} the target
average is over all treated, which we split by exposure,
\begin{align}\label{puresplit}
\tau_D^{1}(0)=q_0\,\Exp{\tau_D^{1}(X,0)\mid \cTc{0}{},T=1}
+\left(1-q_0\right)\vartheta\!\left(\cT\setminus\cTc{0}{},\,\cTc{0}{};V\right),
\end{align}
where $q_0=\Pr(G=0\mid D=1,T=1)$. The first term needs no adjustment. The second
carries the conditional pure direct effect from the non-exposed to the exposed
treated, again as a linear combination of period-specific terms with donor cells
$\{D=1,G=0,T=t\}$ and target $\{D=1,G\neq0,T=1\}$. The statistic underlying the
test of \eqref{R1} is of the same form with target $\{D=1,T=1\}$ and donor cells
$\{D=0,G=0,T=t\}$, and the statistic underlying \eqref{R2} combines the donor cells
$\{D=0,G=g,T=t\}$ and $\{D=1,T=t\}$ for $t\leq0$ with the same target. In a
balanced panel with a unit-level covariate vector each combination reduces to a
single term. The transfer in \eqref{puresplit} becomes
$\vartheta(\cT\setminus\cTc{0}{},\cTc{0}{};S)$ and the statistic underlying the test
of \eqref{R1} becomes $\vartheta(\cT_1,\cN{0}{1};S)$, both with $V=S$. The statistic
underlying \eqref{R2} becomes the difference between $\Exp{V\mid X,\cN{g}{}}$ and
$\Exp{V\mid X,\cT}$ with $V=\sum_{t\leq0}w_tY_t-Y_0$, the extrapolated
pre-treatment trend of a unit.

\section{Proofs}\label{app}

\paragraph{Proof of Proposition~\ref{prop:main}.} By Assumption~\ref{A:noant},
$Y_t=Y_t(0,0)$ for $t\leq0$ in every cell, such that the pre-treatment
conditional means $\mu_t(X,\mathcal{C})$ are conditional means of $Y_t(0,0)$ and
$\nu(X,\mathcal{C})$ is identified. Assumption~\ref{A:ITS} then gives
$\Exp{Y_1(0,0)\mid X,\mathcal{C}}=\nu(X,\mathcal{C})$. With $\mathcal{C}=\{D=1\}$
and $Y_1=Y_1(1,G)$ among the treated,
$\tau_T^{1}(X)=\Exp{Y_1(1,G)\mid X,D=1,T=1}-\Exp{Y_1(0,0)\mid X,D=1,T=1}
=\mu_1(X,\cT)-\nu(X,\cT)$, and with $\mathcal{C}=\{D=0,G=g\}$ and
$Y_1=Y_1(0,g)$ in that cell the same argument gives $\tau_I^{0}(X,g)$. The
aggregates are the averages of the conditional effects over the covariate
distribution of the respective cell in the post-treatment period, by definition in
\eqref{tatt} and \eqref{se}. If that distribution is the same in all periods, then
$\Exp{\mu_t(X,\mathcal{C})\mid\mathcal{C},T=1}=\Exp{Y_t\mid\mathcal{C},T=t}$ for
all $t$ by the law of iterated expectations, such that under \eqref{lintrend} the
aggregates collapse to $\Exp{S\mid\mathcal{C}}$ with
$S=Y_1-\sum_{t\leq0}w_tY_t$, which in a balanced panel is the mean of a
within-unit contrast. \hfill$\square$

\paragraph{Proof of Proposition~\ref{prop:direct}.} \textit{(i)} By \eqref{pure},
$\tau_T^{1}-\tau_D^{1}(G)=\tau_I^{1}(0)=\Exp{Y_1(0,G)-Y_1(0,0)\mid D=1}$, which holds
by the definitions alone. By the law of iterated expectations,
Assumption~\ref{A:transfer} and
$\varsigma(X,g)=\Exp{Y_1(0,g)-Y_1(0,0)\mid X,D=1,G=g}$,
\begin{align*}
\tau_I^{1}(0)
&=\sum_{g\in\mathcal{G}}\Pr(G=g\mid D=1,T=1)\,
\Exp{\varsigma(X,g)\mid D=1,G=g,T=1}\\
&=\sum_{g\in\mathcal{G}}\Pr(G=g\mid D=1,T=1)\,\Exp{\tau_I^{0}(X,g)\mid D=1,G=g,T=1},
\end{align*}
where the common support part of Assumption~\ref{A:transfer} ensures that
$\tau_I^{0}(X,g)$ is well defined at the covariate values and exposure levels
occurring among the treated, and the second equality uses
Proposition~\ref{prop:main}.

\textit{(ii)} Applying Proposition~\ref{prop:main} to the cell $\{D=1,G=0\}$, in
which $Y_1=Y_1(1,0)$, gives
$\mu_1(X,\cTc{0}{})-\nu(X,\cTc{0}{})=\Exp{Y_1(1,0)-Y_1(0,0)\mid X,D=1,G=0}$. Under
Assumption~\ref{A:transfer2} the first term of this difference does not vary with
$G$ within the treated given $X$, and the second does not either, since
$\Exp{Y_1(0,0)\mid X,D=1}$ does not involve $G$, such that it equals
$\tau_D^{1}(X,0)=\Exp{Y_1(1,0)-Y_1(0,0)\mid X,D=1}$. Averaging over the covariate
distribution of the treated in the post-treatment period gives $\tau_D^{1}(0)$, and
$\tau_I^{1}(1)=\tau_T^{1}-\tau_D^{1}(0)$ follows from \eqref{pure}. The common support
part of Assumption~\ref{A:transfer2} ensures that $\tau_D^{1}(X,0)$ is well defined
on the support of $X$ among the treated. \hfill$\square$

\paragraph{Proof of Proposition~\ref{prop:alt}.} In the cell $\{D=0,G=g\}$,
$Y_1=Y_1(0,g)$, such that
\begin{align}\label{contam}
\mu_1(X,\cN{g}{})=\Exp{Y_1(0,0)\mid X,D=0,G=g}+\tau_I^{0}(X,g).
\end{align}
Under Assumptions~\ref{A:CT} and \ref{A:noant},
\begin{align*}
\Exp{Y_1(0,0)\mid X,D=0,G=g}=\mu_0(X,\cN{g}{})+\Exp{Y_1(0,0)-Y_0(0,0)\mid X,D=1},
\end{align*}
and by Assumption~\ref{A:ITS} for $\{D=1\}$ the last term equals
$\nu(X,\cT)-\mu_0(X,\cT)$, which gives \eqref{ctroute}. If
Assumption~\ref{A:ITS} also holds for $\{D=0,G=g\}$, then
$\Exp{Y_1(0,0)\mid X,D=0,G=g}=\nu(X,\cN{g}{})$ and $\tau_I^{0}(X,g)$ is as in
Proposition~\ref{prop:main}; the two expressions coincide if and only if the two
extrapolated conditional trends coincide. \hfill$\square$

\paragraph{Proof of Proposition~\ref{prop:dr}.} Write
$u_1=\I\{\mathcal{A}\}/p$ and
$u_0=\I\{\mathcal{C}\}\pi(X)/\{(1-\pi(X))p\}$, both defined within $R=1$.
Since $\Exp{u_1\mid X}=\Exp{u_0\mid X}=\pi(X)/p$ by Bayes' theorem,
$\Exp{u_1h(X)}=\Exp{u_0h(X)}=\Exp{h(X)\mid\mathcal{A}}$ for any integrable $h$.
Hence $\Exp{u_1m(X)}=\vartheta(\mathcal{A},\mathcal{C};V)$ and, using
$\Exp{V-m(X)\mid X,\mathcal{C}}=0$,
$\Exp{u_0\{V-m(X)\}}=\Exp{\frac{\pi(X)}{p}\Exp{V-m(X)\mid X,\mathcal{C}}}=0$, which
gives $\Exp{\psi}=0$. For Neyman orthogonality, let $\eta_r=\eta_0+r(\eta-\eta_0)$
and let $\partial_r$ denote the derivative at $r=0$, with $h_m$ and $h_\pi$ the
directions of $m$ and $\pi$. The direction $h_m$ contributes
$\Exp{u_1h_m(X)}-\Exp{u_0h_m(X)}=0$ by the display above, and $h_\pi$ contributes
\begin{align*}
\Exp{\frac{\I\{\mathcal{C}\}h_\pi(X)}{\left(1-\pi(X)\right)^2p}
\left\{V-m(X)\right\}}
=\Exp{\frac{h_\pi(X)\Pr(\mathcal{C}\mid X)}{\left(1-\pi(X)\right)^2p}
\cdot\Exp{V-m(X)\mid X,\mathcal{C}}}=0.
\end{align*}
Since the score is linear in $m$ and smooth in $\pi$, which is bounded away from
one, the Gateaux derivative in a general direction is the sum of the two and hence
zero. \hfill$\square$

\paragraph{Proof of Theorem~\ref{thm:1}.} The proof is only sketched, since it closely follows the argument for the average treatment effect on the treated in \citet{Chetal2018}. The score \eqref{score} has the same structure as the doubly robust score for that parameter. If the
covariate composition of a cell is
stable across periods, the estimators of $\tau_T^{1}$ and $\tau_I^{0}(g)$ are sample
means and are $\sqrt{n}$-consistent and asymptotically normal by the central limit
theorem under Assumption~\ref{A:reg}(a); otherwise they are linear combinations of
terms of the form $\vartheta(\mathcal{A},\mathcal{C};V)$ and covered by the argument
below. For $\vartheta(\mathcal{A},\mathcal{C};V)$, the score \eqref{score} is linear in the
parameter with Jacobian $-1$, such that by Proposition~\ref{prop:dr} and
Assumption~\ref{A:reg} the conditions of Theorem 3.1 of \citet{Chetal2018} hold:
the empirical process term vanishes by cross-fitting and consistency of the
nuisance estimators, and the bias term is bounded by the product of their
estimation errors by orthogonality, which is $o_p(n^{-1/2})$ by
Assumption~\ref{A:reg}(d). Stacking the influence functions and applying the
multivariate central limit theorem gives joint asymptotic normality. The results
for $\tau_D^{1}(G)$, $\tau_D^{1}(0)$ and $\tau_I^{1}(1)$ follow from the delta method applied to
\eqref{dirid} and \eqref{puresplit}, whose weights $\Pr(G=g\mid D=1,T=1)$ are
estimated at the parametric rate. \hfill$\square$

\end{document}